\documentclass[sigconf]{acmart}

\makeatletter
\def\subsubsection{\setlength\parindent{10pt}\@startsection{subsubsection}{3}%
  \z@{.5\linespacing\@plus.7\linespacing}{.1\linespacing}%
  {\normalfont\itshape}}
\makeatother

\usepackage{booktabs} 
\usepackage{multirow}
\usepackage{amssymb}
\usepackage{amsmath}
\usepackage{caption}
\usepackage{multicol}
\usepackage{graphicx}
\usepackage{subfig}
\usepackage{enumitem}

\setcopyright{rightsretained}

\def\:{\hskip0pt} 

\acmConference[DESIRES 2018]{Design of Experimental Search \&  Information REtrieval Systems}{August 2018}{Bertinoro, Italy} 
\acmYear{2018}
\copyrightyear{2018}

\author{Hamed Zamani}
\affiliation{%
  \institution{Center for Intelligent Information Retrieval}
  \institution{College of Information and Computer Sciences}
  \institution{University of Massachusetts Amherst}
  \city{Amherst, MA 01003}
}
\email{zamani@cs.umass.edu}

\author{W. Bruce Croft}
\affiliation{%
  \institution{Center for Intelligent Information Retrieval}
  \institution{College of Information and Computer Sciences}
  \institution{University of Massachusetts Amherst}
  \city{Amherst, MA 01003}
}
\email{croft@cs.umass.edu}

\begin{document}
\title{Joint Modeling and Optimization of Search and Recommendation}

\begin{abstract}
Despite the somewhat different techniques used in developing search engines and recommender systems, they both follow the same goal: helping people to get the information they need at the right time. Due to this common goal, search and recommendation models can potentially benefit from each other. The recent advances in neural network technologies make them effective and easily extendable for various tasks, including retrieval and recommendation. This raises the possibility of jointly modeling and optimizing search ranking and recommendation algorithms, with potential benefits to both. In this paper, we present theoretical and practical reasons to motivate joint modeling of search and recommendation as a research direction. We propose a general framework that simultaneously learns a retrieval model and a recommendation model by optimizing a joint loss function. Our preliminary results on a dataset of product data indicate that the proposed joint modeling substantially outperforms the retrieval and recommendation models trained independently. We list a number of future directions for this line of research that can potentially lead to development of state-of-the-art search and recommendation models.
\end{abstract}

\keywords{Information retrieval, information filtering, search engine, recommender system}

\maketitle

\begin{figure}[t]
    \centering
    \includegraphics[scale=0.7]{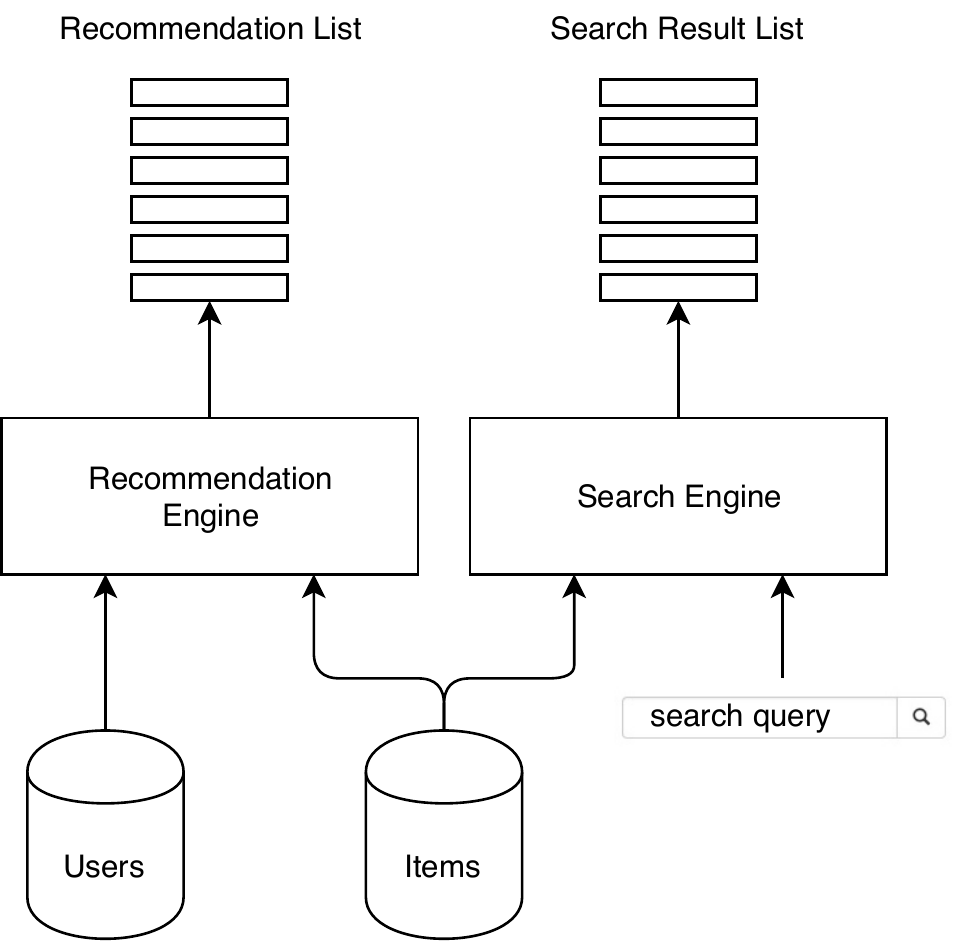}
    \caption{An example of joint search (without personalization) and recommendation systems where items are shared, e.g., in e-commerce websites. The intuition behind joint modeling of search and recommendation is making use of training data from both sides to learn more accurate item representations.}
    \label{fig:jsr}
\end{figure}

\section{Introduction}
\label{sec:intro}
A quarter century has passed since \citet{Belkin:1992} discussed the similarity and unique challenges of information retrieval (IR) and information filtering (IF) systems. They concluded that their underlying goals are essentially equivalent, and thus they are two sides of the same coin. This is why content-based filtering approaches, especially those deal with unstructured data, employ several techniques initially developed for IR tasks, e.g., see \cite{Lavrenko:2000,Lops:2011,Rahmatizadeh:2017,Zamani:2018:IRJ}. With the growth of collaborative filtering approaches, IR and recommender system (RecSys) have become two separate fields with a little overlap between the two communities. Nevertheless, IR models and evaluation methodologies are still common in recommender systems. For instance, common IR evaluation metrics such as mean average precision (MAP) and normalized discounted cumulative gain (NDCG) \cite{Jarvelin:2002} are frequently used by the RecSys community \cite{Schedl:2018}. IR models such as learning to rank approaches are also popular in the RecSys literature \cite{Karatzoglou:2013}. \citet{Costa:2011} formulated recommender systems as an IR task. The language modeling framework for information retrieval \cite{Ponte:1998} and relevance models \cite{Lavrenko:2001} have been also adapted for the collaborative filtering task \cite{Parapar:2013,Wang:2006,Wang:2008}. On the other hand, RecSys techniques have been also used in a number of IR tasks. For instance, \citet{Zamani:2016:CIKM} cast the query expansion task to a recommendation problem, and used a collaborative filtering approach to design a pseudo-relevance feedback model. 

In this paper, we revisit the \citeauthor{Belkin:1992}'s insights to relate these two fields once again. We believe that search engines and recommender systems seek the same goal:

\medskip
\emph{Helping people get the information they need at the right time.}
\medskip

\noindent Therefore, from an abstract point of view, joint modeling and optimization of search engines and recommender systems, if possible, could potentially benefit both systems. Successful implementation of such joint modeling could close the gap between the IR and RecSys communities. Moreover, joint optimization of search and recommendation is an interesting and feasible direction from the application point of view. For example, in e-commerce websites, such as Amazon\footnote{\url{https://www.amazon.com/}} and eBay\footnote{\url{https://www.ebay.com/}}, users use the search functionality to find the products relevant to their information needs, and the recommendation engine recommends them the products that are likely to address their needs. This makes both search and recommendation the two major components in e-commerce websites. As depicted in \figurename~\ref{fig:jsr}, they share the same set of products (and potentially users in case of personalized search), and thus the user interactions with both search engine and recommender system can be used to improve the performance in both retrieval and recommendation. Note that this is not only limited to the e-commerce websites; any service that provides both search and recommendation functionalities can benefit from such joint modeling and optimization. This includes media streaming services, such as Netflix and Spotify, media sharing services, such as YouTube, academic publishers, and news agencies.

Deep learning approaches have recently shown state-of-the-art performance in various retrieval \cite{Dehghani:2017,Guo:2016,Mitra:2017,Zamani:2018:WSDM} and recommendation tasks \cite{Bansal:2016,He:2017}. Recently, \citet{Ai:2017} and \citet{Zhang:2017} showed that using multiple sources of information is useful in both product search and recommendation, which was made possible by neural models in both applications. These neural retrieval and recommendation models can be combined and trained jointly, which is the focus of this paper. We propose a general framework, called JSR,\footnote{JSR stands for the joint search and recommendation framework.} to jointly model and train search engines and recommender systems. As the first step towards implementing the JSR framework, we use simple fully-connected neural networks to investigate the promise of such joint modeling. We evaluate our models using Amazon's product dataset. Our experiments suggest that joint modeling can lead to substantial improvements in both retrieval and recommendation performance, compared to the models trained separately. We show that joint modeling can also lead to higher generalization by preventing the model to overfit on the training data.
The observed substantial improvements suggest this research direction as a new promising avenue in the IR and RecSys literature. We finish by describing potential outcomes for this research direction.

\begin{figure*}[t]
    \centering
    \includegraphics[width=.7\linewidth]{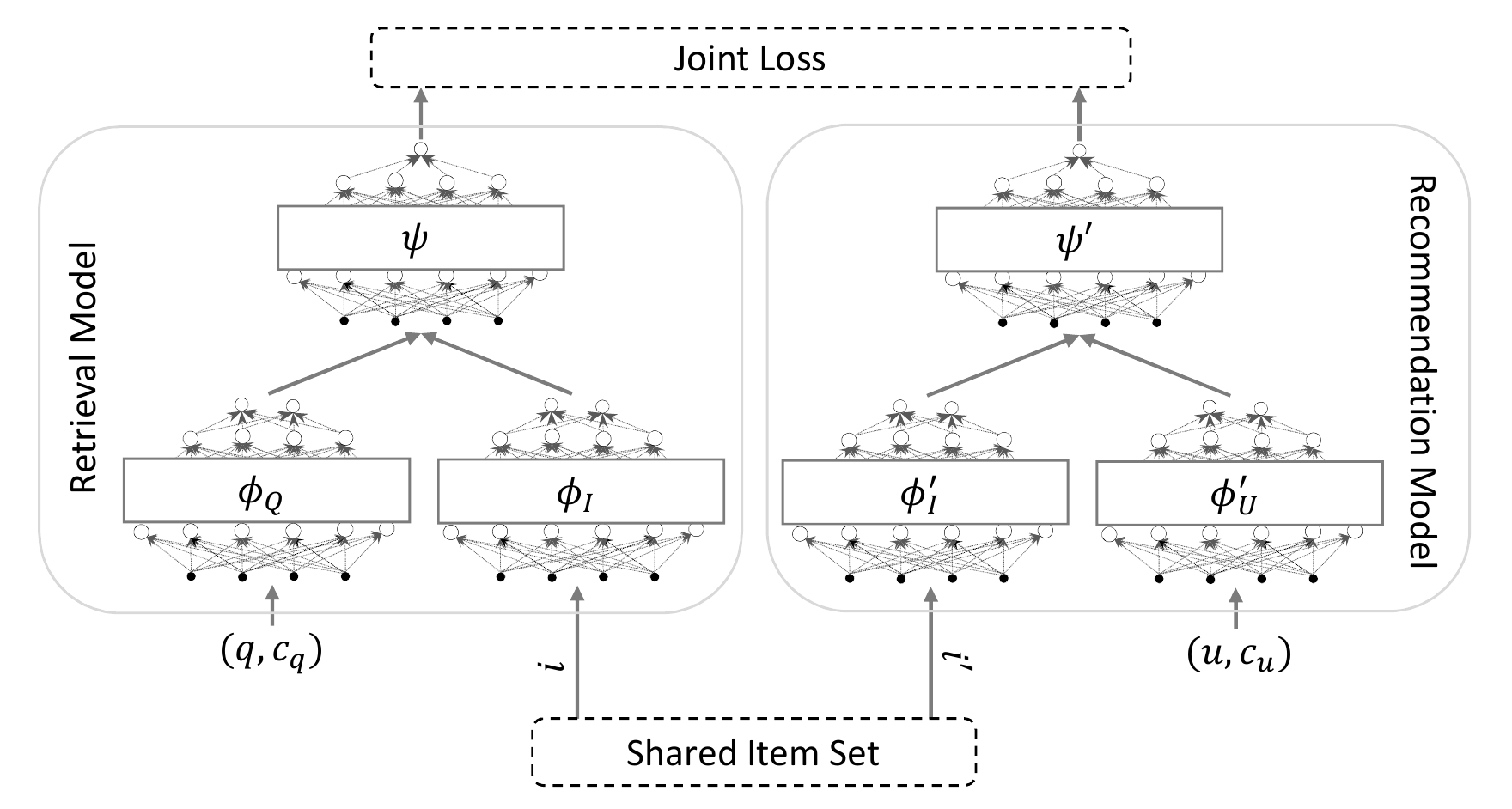}
    \caption{Overview of the JSR Framework. JSR learns a retrieval model and a recommendation model based on a shared set of items and a joint loss function.}
    \label{fig:framework}
\end{figure*}

\section{The Joint Search-Recommendation Framework}
\label{sec:method}
In this section, we describe our simple framework for joint modeling and optimization of search engines and recommender systems, called \emph{JSR}. The purpose of JSR is to take advantage of both search and recommendation training data in order to improve the performance in both tasks. This can be achieved by learning joint representations and simultaneous optimization. In the following subsections, we simplify and formalize the task and further introduce the JSR framework.

\subsection{Problem Statement}
\label{sec:method:problem}
Given a set of retrieval training data (e.g., a set of relevant and non-relevant query-item pairs) and a set of recommendation training data (e.g., a set of user-item-rating triples), the task is to train a retrieval model and a recommender system, jointly. Formally, assume that $I=\{i_1, i_2, \cdots, i_k\}$ is a set of $k$ items. Let $D_{IR} = \{(q_1, R_1, \overline{R}_1), (q_2, R_2, \overline{R}_2), \cdots, (q_n, R_n, \overline{R}_n)\}$ be a set of retrieval data, where $R_i \subseteq I$ and $\overline{R}_i \subseteq I$ respectively denote the set of relevant and non-relevant items for the query $q_i$. Hence, $R_i \cap \overline{R}_i = \emptyset$. Also, let $D_{RS} = \{(u_1, I_1), (u_2, I_2), \cdots, (u_m, I_m)\}$  be a set of recommendation data where $I_i \subseteq I$ denotes the set of items favored (e.g., purchased) by the user $u_i$.\footnote{This can be simply generalized to numeric ratings, as well.} Assume that $D_{IR}$ is split to two disjoint subsets $D_{IR}^{train}$ and $D_{IR}^{test}$ by query, i.e., there is no query overlap between these two subsets. Also, assume that $D_{RS}$ is split to two disjoint subsets $D_{RS}^{train}$ and $D_{RS}^{test}$, such that both subsets include all users and $D_{RS}^{train}$ contains a random subset of purchased items by each user and $D_{RS}^{test}$ contains the remaining items. This means that there is no user-item overlap between $D_{RS}^{train}$ and $D_{RS}^{test}$. Note that although the training data for search ranking differs from the data used for training a recommender system, they both share the same set of items. 

The task is to train a retrieval model $\mathcal{M}_{IR}$ and a recommendation model $\mathcal{M}_{RS}$ on the training sets $D_{IR}^{train}$ and $D_{RS}^{train}$. The models $\mathcal{M}_{IR}$ and $\mathcal{M}_{RS}$ will be respectively evaluated based on the retrieval performance on the test queries in $D_{IR}^{test}$ and the recommendation performance based on predicting the favorite (e.g., purchased) items for each user in the test set $D_{RS}^{test}$. Note that $\mathcal{M}_{IR}$ and $\mathcal{M}_{RS}$ may share some parameters.

\subsection{The JSR Framework}
\label{sec:method:jsr}
JSR is a \emph{general} framework for jointly modeling search and recommendation and consists of two major components: a retrieval component and a recommendation component. The retrieval component computes the retrieval score for an item $i$ given a query $q$ and a query context $c_q$. The query context may include the user profile, long-term search history, session information, or situational context such as location. The recommendation component computes a recommendation score for an item $i$ given a user $u$ and a user context $c_u$. The user context may consist of the recent user's activities, the user's mood, situational context, etc. \figurename~\ref{fig:framework} depicts a high-level overview of the JSR framework. Formally, the JSR framework calculates the following two scores:
\begin{align}
    & \text{retrieval score} = \psi(\phi_Q(q, c_q), \phi_I(i)) \\
    & \text{recommendation score} = \psi'(\phi'_U(u, c_u), \phi'_I(i))
\end{align}
where $\psi$ and $\psi'$ are the matching functions, and $\phi_Q$, $\phi_I$, $\phi'_U$, and $\phi'_I$ are the representation learning functions. In the following subsection, we describe how we implement these functions using fully-connected feed-forward networks. This framework can be further implemented using more sophisticated and state-of-the-art search and recommendation network architectures. Note that the items are shared by both search and recommendation systems, thus they can benefit from an underlying shared representation for each item. For simplicity, we do not consider context in the initial framework described here.

Independent from the way each component is implemented, we train the JSR framework by minimizing a joint loss function $\mathcal{L}$ that is equal to the sum of retrieval loss and recommendation loss, as follows:
\begin{equation}
    \mathcal{L}(b, b') = \mathcal{L}_{IR}(b) + \mathcal{L}_{RS}(b')
\end{equation}
where $b$ and $b'$ are two mini-batches containing training data for search and recommendation, respectively. We train both search and recommendation models using pairwise training. Therefore, each training instance for the retrieval model is a query $q_j$ from $D_{IR}^{train}$, a positive item sampled from $R_j$, and a negative item sampled from $\overline{R}_j$. $\mathcal{L}_{IR}(b)$ is a binary cross-entropy loss function (i.e., equivalent to negative likelihood) as follows:
\begin{align}
    \mathcal{L}_{IR}(b) & = - \sum_{j=1}^{|b|} {\log{p(i_j > \overline{i}_j | q_j)}} \nonumber\\
    & = - \sum_{j=1}^{|b|} {\log{\frac{\exp (\psi(\phi_Q(q_j), \phi_I(i_j)))}{\exp (\psi(\phi_Q(q_j), \phi_I(i_j))) + \exp (\psi(\phi_Q(q_j), \phi_I(\overline{i}_j)))}}}\nonumber
\end{align}

The recommendation loss is also defined similarly; for each user $u_j$, we draw a positive sample $i_j$ from the user's favorite items (i.e., $I_j$ in $D_{RS}^{train}$), and a random negative sample $\overline{i}_j$ from $I$. $\mathcal{L}_{RS}(b)$ is also defined as a binary cross-entropy loss function as follows:
\begin{align}
    \mathcal{L}_{RS}(b') & = - \sum_{j=1}^{|b'|} {\log{p(i_j > \overline{i}_j | u_j)}} \nonumber\\
    & = - \sum_{j=1}^{|b|} {\log{\frac{\exp (\psi'(\phi'_U(u_j), \phi'_I(i_j)))}{\exp (\psi'(\phi'_U(u_j), \phi'_I(i_j))) + \exp (\psi'(\phi'_U(u_j), \phi'_I(\overline{i}_j)))}}}\nonumber
\end{align}

In summary, the search and recommendation components in the JSR framework are modeled as two distinct functions that may share some parameters. They are optimized via a joint loss function that minimizes pairwise error in both retrieval and recommendation, simultaneously.

\begin{table*}
\centering
\caption{Statistics for the three product categories used in our experiments. The data is extracted from Amazon's product data.}
\begin{tabular}{p{5cm}llll} \toprule
Category & \# reviews & \# items & \# users & \# queries \\\hline
Electronics & 1,689,188 & 63,001 & 192,403 & 989 \\
Kindle Store & 989,618 & 61,934 & 68,223 & 4,603 \\
Cell Phones and Accessories & 194,439 & 10,429 & 27,879 & 165 \\\bottomrule
\end{tabular}
\label{tab:dataset}
\end{table*}

\subsection{Implementation of JSR}
\label{sec:method:network}
Since the purpose of this paper is to only show the potential importance of joint modeling and optimization of search and recommendation models, we simply use fully-connected feed-forward networks to implement the components of the JSR framework. The performance of more sophisticated search and recommendation models will be investigated in the future. As mentioned earlier in Section \ref{sec:method:jsr}, we do not consider query and user contexts in our experiments.

We model the query representation function $\phi_Q$ as a fully-conn\-ected network with a single hidden layer. The weighted average of embedding vectors for individual query terms is fed to this network. In other words, $\sum_{t \in q} \widehat{\mathcal{W}}(t)\cdot \mathcal{E}(t)$ is the input of the query representation network, where $\mathcal{W}: V \rightarrow \mathbb{R}$ maps each term in the vocabulary set $V$ to a global real-valued weight and $\mathcal{E}: V \rightarrow \mathbb{R}^d$ maps each term to a $d$-dimensional embedding vector. Note that the matrices $\mathcal{W}$ and $\mathcal{E}$ are optimized as part of the model at the training time. $\widehat{\mathcal{W}}(t)$ is just a normalized weight computed using a softmax function as $\frac{\exp(\mathcal{W}(t))}{\sum_{t' \in q}{\exp(\mathcal{W}(t'))}}$. This simple yet effective bag-of-words representation has been previously used in \cite{Dehghani:2017,Zamani:2018:SIGIR} for the ad-hoc retrieval and query performance prediction tasks. The item representation functions $\phi_I$ and $\phi'_I$ are also implemented similarly. The matrices $\mathcal{W}$ and $\mathcal{E}$ are shared by all of these functions for transferring knowledge among the retrieval and recommendation components.

The user representation function $\phi'_U$ is simply implemented as a look-up table that returns the corresponding row of a user embedding matrix $\mathcal{U}: U \rightarrow \mathbb{R}^{d'}$ that maps each user to a $d'$-dimensional dense vector. The model learns appropriate user representations based on the items they previously rated (or favored) in the training data.

The matching functions $\psi$ and $\psi'$ are implemented as two layer fully-connected networks. The input of $\psi$ is $\phi_Q \circ \phi_I$ where $\circ$ denotes the Hadamard product. Similarly, $\phi'_U \circ \phi'_I$ is fed to the $\psi'$ network. This enforces the outputs of $\phi_Q$ and $\phi_I$ as well as $\phi'_U$ and $\phi'_I$ to have equal dimensionalities. Note that both $\psi$ and $\psi'$ each returns a single real-valued score. These matching functions are similar to those used in \cite{Mitra:2017,Zamani:2018:WSDM} for web search. 

In each network, we use ReLU as the activation function in the hidden layers and sigmoid as the output activation function. We also use dropout in all hidden layers to prevent overfitting.

\section{Preliminary Experiments}
\label{sec:exp}
In this section, we present a set of preliminary results that provide insights into the advantages of jointly modeling and optimizing search engines and recommender systems. Note that to fully understand the value of the proposed framework, large-scale and detailed evaluation and analysis are required and will be done in future work.

In the following, we first introduce our data for training and evaluating both search and recommendation components. We further review our experimental setup and evaluation metrics, which are followed by the preliminary results and analysis.

\subsection{Data}
Experiment design for the search-recommendation joint modeling task is challenging, since there is no public data available for both tasks with a shared set of items. To evaluate our models, we used the Amazon product dataset\footnote{\url{http://jmcauley.ucsd.edu/data/amazon/}} \cite{He:2016,McAuley:2015}, consisting of millions of users and products, as well as rich meta-data information including user reviews, product categories, and product descriptions. The data only contains the users and items with at least five associated reviews. In our experiments, we used three subsets of this dataset associated with the following categories: Electronics, Kindle Store, and Cell Phones \& Accessories. The first two are large-scale datasets covering common product types, while the last one is a small dataset suitable for evaluating the models in a scenario where data is limited.

\textbf{Recommendation Data}: In the Amazon website, users can only submit reviews for the products that they have already purchased. Therefore, from each review we can infer that the user who wrote it has purchased the corresponding item. This results in a set of purchased (user, item) pairs for constructing the set $D_{RS}$ (see Section \ref{sec:method:problem}) that can be used for training and evaluating a recommender system.

\textbf{Retrieval Data}: The Amazon product data does not contain search queries, thus cannot be directly used for evaluating retrieval models. As \citet{Rowley:2000} investigated, directed product search queries contain either a producer's name, a brand, or a set of terms describing the product category. Following this observation, \citet{VanGysel:2016} proposed to automatically generate queries based on the product categories. To be exact, for each item in a category $c$, a query $q$ is generated based on the terms in the category hierarchy of $c$. Then, all the items within that category are marked as relevant for the query $q$. The detailed description of the query generation process can be found in \cite{Ai:2017}. A set of random negative items are also sampled as non-relevant items to construct $D_{IR}$ (see Section~\ref{sec:method:problem}) for training.

\subsection{Experimental Setup}
We cleaned up the data by removing non-alphanumerical characters and stopwords from queries and reviews. Similar to previous work~\cite{Ai:2017}, the content of reviews for each item $i$ were concatenated to represent the item.

We implemented our model using TensorFlow.\footnote{\url{https://www.tensorflow.org/}} In all experiments, the network parameters were optimized using Adam optimizer \cite{Kingma:2014}. Hyper-parameters were optimized using grid search based on the loss value obtained on a validation set (the model was trained on $90\%$ of the training set and the remaining $10\%$ was used for validation). The learning rate was selected from $\{1E-5, 5E-4, 1E-4, 5E-4, 1E-3\}$. The batch sizes for both search and recommendation (see $|b|$ and $|b'|$ in Section~\ref{sec:method:jsr}) were selected from $\{32, 64, 128, 256\}$. The dropout keep probability was selected from $\{0.5, 0.8, 1.0\}$. The word and user embedding dimensionalities were set to $200$ and the word embedding matrix was initialized by the GloVe vectors \cite{Pennington:2014} trained on Wikipedia 2014 and Gigawords 5.\footnote{The pre-trained vectors are accessible via \url{https://nlp.stanford.edu/projects/glove/}.}

\subsection{Evaluation Metrics}
To evaluate the retrieval model, we use mean average precision (MAP) of the top $100$ retrieved items and normalized discounted cumulative gain (NDCG) of the top $10$ retrieved items (NDCG@10). To evaluate the recommendation performance, we use NDCG, hit ratio (Hit), and recall. The cut-off for all recommendation metrics is $10$. Hit ratio is defined as the ratio of users that are recommended at least one relevant item.

\begin{table*}[t]
    \centering
    \caption{Retrieval performance of the model trained independently or jointly with a recommendation model. The superscript $*$ indicates that the improvements are statistically significant, at the 0.05 level using the paired two-tailed t-test.}
    \begin{tabular}{lcccccc} \toprule
        \multirow{2}{*}{Method} & \multicolumn{2}{c}{Electronics} & \multicolumn{2}{c}{Kindle Store} & \multicolumn{2}{c}{Cell Phones} \\\cline{2-7}
         & MAP & NDCG@10 & MAP & NDCG@10 & MAP & NDCG@10 \\\hline
        Individual Training & 0.243 & 0.283 & 0.031 & 0.028 & 0.073 & 0.086 \\
        Joint Training & \textbf{0.317}* & \textbf{0.388}* & \textbf{0.149}* & \textbf{0.126}* & \textbf{0.130}* & \textbf{0.204}* \\ \bottomrule
    \end{tabular}
    \label{tab:ir}
\end{table*}

\begin{table*}[t]
    \centering
    \caption{Recommendation performance of the model trained independently or jointly with a retrieval model. The superscript $*$ indicates that the improvements are statistically significant, at the 0.05 level using the paired two-tailed t-test.}
    \begin{tabular}{lccccccccc} \toprule
        \multirow{2}{*}{Method} & \multicolumn{3}{c}{Electronics} & \multicolumn{3}{c}{Kindle Store} & \multicolumn{3}{c}{Cell Phones} \\\cline{2-10}
         & NDCG & Hit & Recall & NDCG & Hit & Recall & NDCG & Hit & Recall \\\hline
        Individual Training & 0.143 & 0.318 & 0.075 & 0.047 & 0.136 & 0.021 & 0.038 & 0.108 & 0.014 \\
        Joint Training & \textbf{0.197}* & \textbf{0.343}* & \textbf{0.092}* & \textbf{0.063}* & \textbf{0.187}* & \textbf{0.034}* & \textbf{0.062}* & \textbf{0.160}* & \textbf{0.034}* \\ \bottomrule
    \end{tabular}
    \label{tab:recsys}
\end{table*}

\subsection{Results and Discussion}
\tablename~\ref{tab:ir} reports the retrieval performance for an individual retrieval model and the one jointly learned with a recommendation model. The results on three categories of the Amazon product dataset demonstrate that the jointly learned model significantly outperforms the individually trained model, in all cases. Note that the network architecture in both models is the same and the only difference is the way that they were trained, i.e., individual training vs. co-training with the recommendation component. We followed the same procedure to optimize the hyper-parameters for both models to have a fair comparison.

The results reported in \tablename~\ref{tab:recsys} also show that the recommendation model jointly learned with a retrieval model significantly outperforms the one trained individually with the same recommendation training data.

In summary, joint modeling and optimization of search and recommendation offers substantial improvements in both search ranking and recommendation tasks. This indicates the potential in joint modeling of these two highly correlated applications.

\begin{figure}[t]
    \centering
    \includegraphics[width=\linewidth]{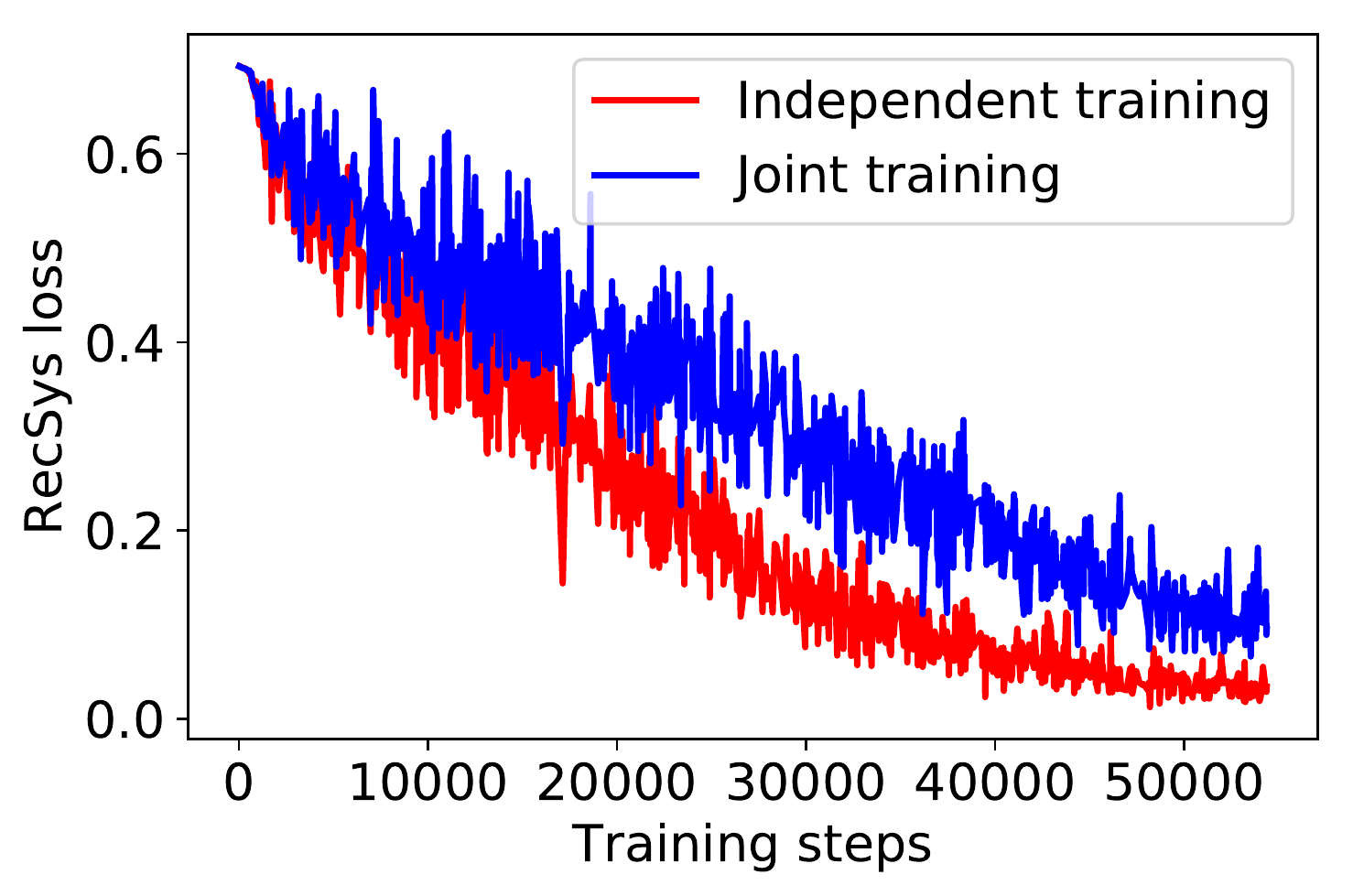}
    \caption{The loss curves for both independent and joint training of the recommendation model on the training data.}
    \label{fig:loss}
\end{figure}

It is important to fully understand the reasons behind such improvements. To this aim, \figurename~\ref{fig:loss} plots the recommendation loss curves on the Cell Phones \& Accessories training data for two recommendation models, one trained individually and the other one trained jointly with the retrieval model. Although the individually learned model underperforms the joint model (see \tablename~\ref{tab:recsys}), its recommendation loss on the training data is less (see \figurename~\ref{fig:loss}). Similar observation can be made from the retrieval loss curves, which are omitted due to the space constraints. It can be inferred that the individually learned model overfits on the training data. Therefore, joint training can be also used as a means to improve generalization by prevention from overfitting.

\noindent \textbf{Example.} Here, we provide an example to intuitively justify the superior performance of the proposed joint modeling. Assume that a query ``iphone accessories'' is submitted. Relevant products include various types iPhone accessories including headphones, phone cases, screen protectors, etc. However, the description and the reviews of most of these items do not match with the term ``accessories''. This results in poor retrieval performance for a retrieval model trained individually. On the other hand, from the recommendation training data, users who bought iPhones, they also bought different types of iPhone accessories. Therefore, the representations learned for these items, e.g., headphones, phone cases, and screen protectors, are close in a jointly trained model. Thus, the retrieval performance for the query ``iphone accessories'' improves, when joint training is employed.

The recommender system can also benefit from the joint modeling. For example, to a user who bought a cell phone, few headphones that have been previously purchased together with this phone by other users have been recommended. From the retrieval training data, all the headphones are relevant to the query ``headphones'' and thus, close representations are learned for all the headphones. This results in recommending the headphones that have not been necessarily purchased by the users together with that phone. This results in substantial improvements in the recall and the overall performance achieved by the recommendation model.

\section{Conclusions and Future Directions}
\label{sec:conclusion}
In this paper, we introduced the search-recommendation joint modeling task by providing intuitions on why jointly modeling and optimizing search engines and recommender systems could be useful in practical scenarios. We performed a set of preliminary experiments to investigate the feasibility of the task and observed substantial improvements compared to the baselines. Our experiments also verified that joint modeling can be seen as a means to improve generalization by prevention from overfitting. This work smooths the path towards studying such a challenging task in practical situations in the future.
 
In the following, we present our insights into the search-recom\-mendation joint modeling task and how it can influence search engines and recommender systems in the future.

An immediate next step should be evaluating the JSR framework in a real-world setting, where queries were issued by real users and different relevance and recommendation signals (e.g., search logs and purchase history) are available for training and evaluation. This would guarantee the actual advantages of the proposed JSR framework in real systems.

Furthermore, given the importance of learning from limited data to both academia and industry \cite{Zamani:2018:SIGIR:LND4IR}, we believe that the significance of JSR could be even greater when training data for either search or recommendation is limited. For instance, assume that an information system has run a search engine for a while and gathered a large amount of user interactions with the system, and a recommender systems has recently been added. In this case, the JSR framework could be particularly useful for transferring the information captured by the search logs to improve the recommendation performance in such a cold-start setting. Even a more extreme case would be of interest where training data for either search or recommendation is available, but no labeled data is in hand for the other task. On the one hand, this extreme case has several practical advantages and enables information systems to provide both search and recommendation functionalities when training data for only one of these functionalities is available. On the other hand, this is a theoretically interesting task, because this is not a typical transfer learning problem; in transfer learning approaches, the distribution of labeled data is often mapped to the distribution of unlabeled target data, which cannot be applied here, since these are two different problems with different inputs. From a theoretical point of view, this extreme case can be viewed as a generalized version of typical transfer learning.

Moreover, in the JSR framework, the search and recommendation components are learned simultaneously. Therefore, improving one of these models (either search or recommendation) can intuitively improve the quality of learned representations. Therefore, this can directly affect the performance of the other task. For example, improving the network architecture for the retrieval model can potentially lead to improvements in the recommendation performance. If future work verifies the correctness of this intuition, this results in ``killing two birds with one stone''.

\section{Acknowledgements}
This work was supported in part by the Center for Intelligent Information Retrieval. Any opinions, findings and conclusions or recommendations expressed in this material are those of the authors and do not necessarily reflect those of the sponsor. The authors thank Qingyao Ai, John Foley, Helia Hashemi, and Ali Montazeralghaem for their insightful comments.  



\end{document}